\documentclass{birkjour}

\usepackage{amsmath}
\usepackage{amssymb}

\usepackage{url}

\usepackage{yhmath}
\usepackage[dvips]{color}

\usepackage{srctex} 

\def\m#1{\mathsf{#1}} 

\newcommand{\cl}[2]{\ensuremath{\mathit{Cl}_{#1,#2}}}

\DeclareMathOperator{\Det}{Det} 

\newcommand{\bbN}{\ensuremath{\mathbb{N}}}

\newcommand{\reverse}[1]{\widetilde{#1}}
\newcommand{\gradeinverse}[1]{\wideparen{#1}}
\newcommand{\cliffordconjugate}[1]{\widetilde{\wideparen{#1}}}

\newcommand{\ee}{\mathrm{e}} 
\newcommand{\dd}{\mathrm{d}} 

\def\A{\mathsf{A}}

\def\m#1{\mathsf{#1}}

\def\e#1{\mathbf{e}_{#1}} 

\newcommand{\ba}{\ensuremath{\mathbf{a}}}
\newcommand{\bb}{\ensuremath{\mathbf{b}}}

\newcommand{\cA}{\ensuremath{\mathcal{A}}}
\newcommand{\cB}{\ensuremath{\mathcal{B}}}

\newcommand{\w}{\ensuremath{\mathbin{\wedge}}} 

\usepackage{color}

\newcommand{\mycomment}[1]{} 

 \theoremstyle{definition}
 
 \theoremstyle{remark}

 \numberwithin{equation}{section}

\newcommand{\magnitude}[1]{\lvert #1\rvert}

\begin{document}


\title[Coordinate-free  exponentials of general MV]
{Coordinate-free  exponentials of general multivector (MV) in
\cl{p}{q} algebras for $p+q=3$}

\author{A.~Acus}
\address{Institute of Theoretical Physics and Astronomy,\br
Vilnius University,\br Saul{\.e}tekio 3, LT-10257 Vilnius,
Lithuania} \email{arturas.acus@tfai.vu.lt}
\thanks{$\dagger$ Corresponding author: A.~Acus}

\author{A.~Dargys}
\address{%
Center for Physical Sciences and Technology,
\br Saul{\.e}tekio 3, LT-10257 Vilnius,
Lithuania} \email{adolfas.dargys@ftmc.lt}

\subjclass{Primary 15A18; Secondary 15A66}

\keywords{ {C}lifford (geometric) algebra, exponentials of
Clifford numbers,  computer-aided theory}


\begin{abstract}
Closed form expressions  in a coordinate-free form in real
Clifford geometric algebras (GAs) $\cl{0}{3}$,  $\cl{3}{0}$,
$\cl{1}{2}$ and $\cl{2}{1}$ are found for  exponential function
when the exponent is the most general multivector (MV). The main
difficulty in solving the problem is connected with an
entanglement or  mixing of vector and bivector components. After
disentanglement, the obtained formulas simplify to the well-known
Moivre-type trigonometric/hyperbolic function for vector or
bivector exponentials. The presented formulas may find wide
application in solving GA differential equations,  in signal
processing, automatic control and robotics.
\end{abstract}

\maketitle





\section{Introduction}
\label{sec:intro}

In the complex number algebra, which is isomorphic to \cl{0}{1}
Clifford geometric algebra (GA),  the exponential can be expanded
into trigonometric function sum (de~Moivre's theorem). In 3D GA
algebras similar formulas are known in special cases only. In
particular, when the square of the blade is equal to $\pm 1$, the
GA exponential can be also expanded in de~Moivre-type sum, in
either trigonometric or hyperbolic functions respectively
\cite{Gurlebeck1997,Lounesto97}. However, expansion of GA
exponential  in case of 3D algebras \cl{3}{0}, \cl{1}{2},
\cl{2}{1} and \cl{0}{3}, when the exponent is a general
multivector (MV), as we shall see is more complicated and as far
as the authors know has not been not analyzed  fully as yet.
Authors of papers~\cite{Chappell2015,Josipovic2019} have
considered general properties of the functions of MV variable for
Clifford algebras $n=p+q\le 3$, including the exponential
function. For this purpose they have  used the property  that the
pseudoscalar $I$ in these algebras commutes with all MV elements
and $I^2=\pm1$. This has allowed to introduce more general
functions related to a polar decomposition of MVs. However, the
analysis is not full enough. A different approach to resolve the
problem is to factor, if possible, the exponential into product of
simpler exponentials, for example, in the  polar
form~\cite{Hitzer2020p,Hitzer2020a,Hitzer2019a,Hitzer2021}.
General bivector exponentials in \cl{4}{1} algebra were analyzed
in~\cite{Cameron2004} in connection with 3D conformal GA.
In~\cite{Dargys2021}, exact closed form  expressions for
coefficients at basis elements to calculate GA exponentials in
coordinate form are presented for all 3D GAs. The final formulas
written in some orthogonal basis appeared rather complicated and
inconvenient in carrying a detailed analysis of the exponentials,
although they may be useful in some cases, for example, for
all-purpose computer programs to calculate GA exponentials with
numerical coefficients.

In this paper the exact exponential formulas~\cite{Dargys2021} are
transformed to coordinate-free form what brings in a clear
geometric interpretation to the problem and to carry a detailed
analysis. Also, special cases when various conditions and
relations are imposed upon  GA elements are considered that may be
useful in applications of formulas in the paper. In
Sec.~\ref{sec:notations} the notation is introduced. In
Sec.~\ref{sec:Cl03} the exponential of the simplest, namely
\cl{0}{3} algebra is considered. Since algebras \cl{0}{3} and
\cl{1}{2} are isomorphic, in Sec.~\ref{sec:Cl30} the exponentials
in the both algebras are investigated simultaneously. In
Sec.~\ref{sec:Cl21} the exponential of \cl{2}{1} is presented. In
Sec.~\ref{sec:Application} possible application of exponentials to
solve GA linear differential equations are presented.  Finally, in
Sec.~\ref{sec:discussion} we discuss further development of the
problem, including the inverse function, viz.  the GA logarithm.


\section{Notation and general properties of GA exponential}
\label{sec:notations} The calculation have been done with GA
program written for \textit{Mathematica}
package~\cite{AcusDargys2017}.  In GA space endowed with
orthonormal basis we expand a general 3D MV in inverse degree
lexicographic ordering: $\{1,\e{1},\e{2},\e{3},\e{12},\e{13},
\e{23}, \\ \e{123}\equiv I\}$, where $\e{i}$ are basis vectors,
$\e{ij}$ are the bivectors and $I$ is the
pseudoscalar.\footnote{\label{note1}An increasing order of digits
in basis elements is used, i.e., we write $\e{13}$ instead of
$\e{31}=-\e{13}$. This convention is reflected in opposite signs
of some terms in the obtained formulas.} The number of subscripts
indicates the grade. The scalar is a grade-0 element, the vectors
$\e{i}$ are the grade-1 elements, etc. In the orthonormalized
basis used here the geometric product of basis vectors satisfy the
anti-commutation relation,
 \begin{equation}\label{anticom}
 \e{i}\e{j}+\e{j}\e{i}=\pm 2\delta_{ij}.
 \end{equation}
For \cl{3}{0} and \cl{0}{3} algebras the squares of basis vectors,
correspondingly, are $\e{i}^2=+1$ and $\e{i}^2=-1$, where
$i=1,2,3$. For mixed signature algebras such as  \cl{2}{1} and
\cl{1}{2} the squares are $\e{1}^2=\e{2}^2=1$, $\e{3}^2=-1$
and $\e{1}^2=1$, $\e{2}^2=\e{3}^2=-1$, respectively.

The general MV that belongs to  real Clifford algebras $\cl{p}{q}$
for $n=p+q= 3$ can be expressed as
\begin{equation}\begin{split}\label{mvA}
\A=&\,a_0+a_1\e{1}+a_2\e{2}+a_3\e{3}+a_{12}\e{12}+a_{23}\e{23}+a_{13}\e{13}+a_{123}I\\
\equiv&\,a_0+\ba+\cA+a_{123}I\equiv a_0+A+a_{123}I,
\end{split}
\end{equation} where $a_i$, $a_{ij}$ and $a_{123}$ are the real
coefficients, and $\ba=a_1\e{1}+a_2\e{2}+a_3\e{3}$ and
$\cA=a_{12}\e{12}+a_{23}\e{23}+a_{13}\e{13}$ is, respectively, the
vector and bivector. $I$~is the pseudoscalar, $I=\e{123}$. In
accord with  \cite{Dargys2021} the sum of vector and bivector is
denoted by symbols $A\equiv\A_{12}=\ba+\cA$. Similarly, the
exponential of $\m{A}$ is denoted as
\begin{equation}
\begin{split}\label{mvB}
\ee^{\m{A}}=&\,\m{B}=b_0+b_1\e{1}+b_2\e{2}+ba_3\e{3}+b_{12}\e{12}+b_{23}\e{23}+ba_{13}\e{13}+a_{123}I\\
\equiv&\,b_0+\bb+\cB+b_{123}I\equiv\,b_0+B+b_{123}I.
\end{split}
\end{equation}

The main involutions, namely the reversion, grade inversion and
Clifford conjugation are denoted, respectively, by tilde,
circumflex and their combination,
\begin{equation}\begin{split}
&\widetilde{\m{A}}=a_0+\ba-\cA-a_{123}I,\\
&\gradeinverse{\m{A}}=a_0-\ba+\cA-a_{123}I,\\
&\cliffordconjugate{\m{A}}=a_0-\ba-\cA+a_{123}I.
\end{split}
\end{equation}

\subsection{\label{expLogTan}General properties of GA  exponential}
The exponential of MV is another MV that  belongs to the same
geometric algebra. Therefore, we shall assume that the defining
equation for exponential is $\ee^{\m{A}}=\m{B}$, where
$\m{A},\m{B}\in\cl{p}{q}$ and $p+q=3$. The following properties
hold for MV exponential:
\begin{equation}\label{expProperties}
\begin{split}
&\exp(\m{A}\m{B})=\exp(\m{A})\exp(\m{B})\quad\text{if and only if\ } \m{A}\m{B}=\m{B}\m{A},\\
&\widetilde{\ee^{\m{A}}}=\ee^{\widetilde{\m{A}}},\quad\gradeinverse{\ee^{\m{A}}}=\ee^{\gradeinverse{\m{A}}},
\quad \cliffordconjugate{\ee^{\m{A}}}=\ee^{\cliffordconjugate{\m{A}}},\\
  &\m{V}\,\exp(\m{A})\m{V}^{-1}=\exp(\m{V}\m{A}\m{V}^{-1}).
\end{split}
\end{equation}
From first formula an exponential identity
$\exp\m{A}=(\exp\m{A/m})^m$, $m\in\bbN$ follows. In the last expression
the transformation $\m{V}$, for example the rotor, has been lifted
to exponent.

The GA exponential $\ee^{\m{A}}$ can be expanded in a series that
has exactly the same structure as a scalar
exponential~\cite{Lounesto97}, from which GA trigonometric and
hyperbolic GA functions  as well as  various other relations that
are analogues of respective scalars  functions
follow~\cite{Dargys2021}. For example,
\begin{equation}\begin{split}
&\cos^2\m{A}+\sin^2\m{A}=1,\quad
\cosh^2\m{A}-\sinh^2\m{A}=1,\\
&\sin(2\m{A})=2\sin\m{A}\cos\m{A}=2\cos\m{A}\sin\m{A},\\
&\cos(2\m{A})=\cos^2\m{A}-\sin^2\m{A}.
\end{split}\end{equation}
Also, it should be noted that GA functions of the same argument
commute. Thus, the sine and cosine functions as well as hyperbolic
GA sine and cosine functions satisfy:
$\sin\m{A}\cos\m{A}=\cos\m{A}\sin\m{A}$ and
$\sinh\m{A}\cosh\m{A}=\cosh\m{A}\sinh\m{A}$.

In sections ~\ref{sec:Cl03}-\ref{sec:Cl21} the exact formulas for
GA exponentials are presented. If the MV is in a numerical form, a
finite  series expansion may be useful as well to know or just to
check the GA formula with the exponential. To minimize the number
of multiplications it is convenient the exponential to represent
in a nested form (aka Horner's rule)
\begin{equation}\label{expHorner}
\ee^{\m{A}}=1+\frac{\m{A}}{1}(1+\frac{\m{A}}{2}(1+\frac{\m{A}}{3}(1+\frac{\m{A}}{4}(1+\dots)))),
\end{equation}
which requires fewer MV products to calculate the truncated series
than working out each power of $\m{A}$. If numerical coefficients
in~$\m{A}$ are not too large  the exponential $\ee^{\m{A}}$ can be
approximated to high precision by~\eqref{expHorner}. The series
may be programmed as a simple iterative procedure repeated
$k$-times that begins from the end (dots) with the initial value
at $\m{A}/k=1$ and then iteratively moving to left.\footnote{In
Mathematica the algorithm reads: expMVHorner$[\m{A}_{-},n_{-}]:=$
$\text{Module}[\{\m{B}=1,s=n+1\},\text{While}[(s=s-1)>0,$
$\m{B}=1.+\text{GP}[\m{B},\m{A}/s]];\m{B}]$, where $n$ is the
number of iterations and GP is the geometric product. If  MV
coefficients are large ($a_J\geq 3$), in addition,  the formula
$(\exp(\m{A}/m))^m$, where $m$ is the integer, may be applied at
first to accelerate the convergence and then to raise the result
to the $m$th power.}

We start from the $\cl{0}{3}$ geometric algebra (GA) where the
expanded exponential in the coordinate form has the simplest MV
coefficients.

\section{MV exponentials in \cl{0}{3} algebra}
\label{sec:Cl03}
\subsection{Exponential in coordinate-free form} Symbolic formulas
in  GA may be written in  coordinate and coordinate-free forms.
The latter presentation is compact and  carries clear geometrical
interpretation and therefore is preferred. However, the formulas
written in the coordinate form sometimes may be  helpful too, in
particular, in GA numerical  calculations with non-symbolic
programmes. In \cite{Dargys2021} we have found  a  general MV
exponentials in coordinate form in all three-dimensional (3D) GAs.
Although the expressions are rather involved, however, they
acquire a simple form if coordinate-free notation (second line in
Eqs.~\eqref{mvA} and ~\eqref{mvB}) is used. Moreover, geometrical
analysis of GA formulas becomes simpler and more evident  when
formulas are rewritten in  a coordinate-free form.

In a case of $\cl{0}{3}$ algebra, after multiplication of
coordinates by respective basis elements and then collection to
vector, bivector, trivector and their products one can transform
the exponential components in~\cite{Dargys2021} to a generic
coordinate-free form,
\begin{align}\label{exp03free}
\exp(\m{A})=
  &\frac{1}{2}\ee^{a_0}
  \Bigl(
  \ee^{a_{123}} (1 + I)  \bigl( \cos a_{+}+\frac{\sin a_{+}}{a_{+}} (\ba + \cA)
     \bigr)\\
      &\qquad + \ee^{-a_{123}}(1 -I ) \bigl(\cos a_{-}+\frac{\sin a_{-}}{a_{-}}  (\ba + \cA)\bigr)\Bigr),\notag\
\end{align}
where $a_{-}$ and $a_{+}$ are the  scalars,
\begin{align}
  a_{-}=&\sqrt{-(\ba\cdot\ba+\cA\cdot\cA) + 2I  \ba\w\cA}\label{aplius}\\
  =&\sqrt{(a_{3}+a_{12})^2+(a_{2}-a_{13})^2+(a_{1}+a_{23})^2},\notag\\
  a_{+}=&\sqrt{-(\ba\cdot\ba+\cA\cdot\cA) - 2I  \ba\w\cA}\label{aminus}\\
  =&\sqrt{(a_{3}-a_{12})^2+(a_{2}+a_{13})^2+(a_{1}-a_{23})^2}\,,\notag
\end{align}
They  show  how the vector and bivector components are mixed up.
The appearance of trigonometric functions in Eq.~\eqref{exp03free}
indicates that the exponential  in \cl{0}{3} has an oscillatory
character. When the denominator in the formula~\eqref{exp03free},
either $a_{+}$ or $a_{- }$, reduces to zero we will have a special
case. The generic formula~\eqref{exp03free} then should be
modified by replacing the corresponding ratios by their limits,
$\lim_{a_{\pm}\to 0}\frac{\sin a_{\pm}}{a_{\pm}}=1$.

If either vector $\ba$ or bivector $\cA$ in
\eqref{exp03free}-\eqref{aminus} is absent then $a_{+}=a_{-}=a$,
where $a$ is a magnitude of the vector
$a=\magnitude{\ba}=\bigl(\ba\cliffordconjugate{\ba}\bigr)^{\frac{1}{2}}=\sqrt{a_1^2+a_2^2+a_3^2}\,$,
or of the bivector
$a=\magnitude{\cA}=\bigl(\cA\cliffordconjugate{\cA}\bigr)^{\frac{1}{2}}=\sqrt{a_{12}^2+a_{13}^2+a_{23}^2}$\,.
If, in addition, the scalar and pseudoscalar are absent,
$a_0=a_{123}=0$, the formula~\eqref{exp03free} reduces to the
well-known trigonometric expressions for exponential of vector and
bivector in a polar form~\cite{Lounesto97},
\begin{equation}\label{MoivreA}
\ee^{\ba}=\cos\magnitude{\ba}+\frac{\ba}{\magnitude{\ba}}\sin\magnitude{\ba},\quad
\ee^{\cA}=\cos\magnitude{\cA}+\frac{\cA}{\magnitude{\cA}}\sin\magnitude{\cA}.
\end{equation}
The trigonometric functions appear in~\eqref{MoivreA} because both
the vector and the  bivector in \cl{0}{3} satisfy $\ba^2<0$ and
$\cA^2<0$. If exponential consists of scalar and pseudoscalar only
then $a_{+}=a_{-}=0$ and the Eq.~~\eqref{exp03free} simplifies to
hyperbolic functions
\begin{equation}\label{MoivreB}
\ee^{a_0+I a_{123}}= \ee^{a_0}(\cosh a_{123}+I\sinh a_{123}),\quad
I^2=1.
\end{equation}

In the following we shall distinguish two kinds of coordinate-free
formulas for exponential functions, namely, generic and special.
The formula \eqref{exp03free} is an example of generic formula
since it is valid for almost all real coefficient $a_J$ values,
where $J$ is a compound index: $J=i$, $ij$, or $ijk$. The
expression~\eqref{MoivreB} represents the special formula, since
in the case $a_{+}=0$ and/or $a_{-}=0$, and as a result we have
division by zero in \eqref{exp03free}   and therefore should use a
modified formula (which, in this case can be obtained by computing
limit of \eqref{exp03free} when $a_{+}\to0$, and/or $a_{-}\to0$).
On the other hand the Eq.~\eqref{MoivreA} represents an important
in practice case of generic solution (obtained by simply equating
the coefficients at scalar and pseudoscalar and, respectively, at
bivector and vector, by zero). For completeness, in a case of
logarithmic functions it would  be interesting to remark that one
may add an additional free MV to the generic or special symbolic
solution~\cite{Acus2021b}. A formula with a free parameters
included could be referred to as general solution. The latter kind
of solution with a free MV is absent for an exponential function.

\vspace{3mm}
 \textbf{Example 1.} {\it Exponential of  MV in \cl{0}{3}.}
Let's compute the  exponential of  $\m{A}=-8-6 \e{2}-9 \e{3}+5
\e{12}-5 \e{13}+6 \e{23}-4 \e{123}$ using the coordinate-free
expression~\eqref{exp03free}. We find $a_{-}=\sqrt{53}$ and
$a_{+}=\sqrt{353}\,$. The exact numerical answer then is
\begin{align}
  \exp(\m{A})= &\tfrac{\ee^{-8}}{2}\biggl(\ee^4 (1-\e{123})\notag\\
  &\hphantom{\frac{\ee^{-8}}{2}\ee^4}
  \times\Bigl(\cos \sqrt{53}+\tfrac{\sin \sqrt{53}}{\sqrt{53}} (-6 \e{2}-9 \e{3}+5 \e{12}-5 \e{13}+6 \e{23})\Bigr)\notag\\
  &\hphantom{\frac{\ee^{-8}}{2}}
  +\ee^{-4}(1+\e{123})\notag\\
  &\hphantom{\frac{\ee^{-8}}{2}}
  \times\Bigl(\cos \sqrt{353}+\tfrac{\sin \sqrt{353}}{\sqrt{353}} (-6 \e{2}-9 \e{3}+5 \e{12}-5 \e{13}+6 \e{23})\Bigr)\biggr).\notag
\end{align}
For comparison, the calculation  of the exponential
series~\eqref{expHorner}  by Mathematica v.12 in a floating point
regime gives six significant figures of the exact solution after
summation of 70 series terms or iterations. For larger iterations
the number of significant figures increases. However it should be
noted that the convergence is not monotonic and to get the first
significant figure for all basis MV elements no less than 64
series terms (iterations) are needed in this particular case.

\subsection{\label{entanglement}Vector-bivector mixing in \cl{0}{3}}
In equations \eqref{aplius} and \eqref{aminus}, the outer product
$\ba\w\cA$ in general case is a trivector. It entangles or mixes
up vector and bivector components in the
exponential~\eqref{exp03free}. This is easy to see if we equate to
zero either $\ba$  or $\cA$. Then, $a_{+}=a_{-}=\magnitude{\ba}$
if $\cA=0$ and $a_{+}=a_{-}=\magnitude{\cA}$ if $\ba=0$, where
$\magnitude{\ba}=\bigl(\ba\cliffordconjugate{\ba}\bigr)^{\frac{1}{2}}$
and
$\magnitude{\cA}=\bigl(\cA\cliffordconjugate{\cA}\bigr)^{\frac{1}{2}}$.
The trivector also vanishes if  $\ba$ and $\cA$ are unequal to
zero but the vector $\ba$ lies in the plane $\cA$. In this
case\footnote{Such a situation is encountered in classical
electrodynamics where magnetic field bivector and electric field
vector lie in a the same plane.} the components satisfy the
condition $I\ba\w\cA=a_1a_{23}-a_2a_{13}+a_3a_{12}=0$. Then the
entanglement (mixing) coefficients become $a_{+}=a_{-}\to
a_m=\sqrt{\magnitude{\ba}^2+\magnitude{\cA}^2}$ and the
exponential~\eqref{exp03free} reduces to
\begin{equation}\label{entangleA}
\ee^{\m{A}}=\ee^{a_{0}}\ee^{Ia_{123}}\big(\cos
a_m+\frac{\ba+\cA}{a_m}\sin a_m\big),\quad\ba\|\cA.
\end{equation}
Thus, the exponential $\ee^{\m{A}}$ in
this case can be factorized. It is interesting that the last
multiplier represents an entangled (mixed up)  Moivre-type formula
for a sum of vector and bivector, where the magnitude of
$(\ba+\cA)$ is
\begin{equation}
a_m
=\magnitude{\ba+\cA}=\sqrt{(\ba+\cA)\cliffordconjugate{(\ba+\cA)}}=\sqrt{\magnitude{\ba}^2+\magnitude{\cA}^2}.
\end{equation}
We shall remind that in Eq.~\eqref{entangleA} the vector $\ba$
lies in the plane~$\cA$. A similar formula can be obtained in
opposite case  if we assumes that, for example, the vector
$\ba\|\e{3}$ and $\cB\|\e{12}$,  i.e., the vector and bivector are
characterized by a single scalar coefficient.\footnote{This
approach  reminds a popular method in physics  where a judicious
choice of mutual orientation of the fields and coordinate vectors
allows to simplify the problem  substantially.} Then the
expression~\eqref{exp03free} gives
\begin{equation}\label{entangleB}
\ee^{\m{A}}=-\big(\cos(a_{12}-a_3)+\frac{a_{12}\e{12}+a_3\e{3}}{a_{12}-a_3}\sin(a_{12}-a_3)\big).
\end{equation}
In conclusion, apart from the pure Moivre-type expressions (see
Eqs.~\eqref{MoivreA} and \eqref{MoivreB}), the generic GA
exponential~\eqref{exp03free} also contains  mixed or entangled
MVs which may be disentangled as shown by Eqs~\eqref{entangleA}
and \eqref{entangleB}) if additional the conditions are imposed.

\section{\label{sec:Cl30}MV exponentials in \cl{3}{0} and \cl{1}{2} algebras}

After multiplication of the scalar coefficients the full
expressions of which are given in \cite{Dargys2021} by respective
basis elements and collection into sum, and finally combining the
resulting expression  into a coordinate-free form we find the
following generic exponential of MV $\m{A}$,
\begin{align}\label{exp30free}
\exp(\m{A})=
  &\ee^{a_0}(\cos a_{123}+I\sin a_{123}) \Bigl(\cos a_{-} \cosh a_{+} +I \sin a_{-} \sinh a_{+}\notag\\
  &+\frac{1}{a_{\smash{+}}^2+a_{\smash{-}}^2} (\cosh a_{+} \sin a_{-}  -I\cos a_{-} \sinh a_{+})\\
  &\hphantom{+\frac{1}{\sqrt{a_{\smash{+}}^2+a_{\smash{-}}^2}}\quad}
  \times\bigl(a_{-}(\ba+\cA) + a_{+} I (\ba+\cA)\bigr) \Bigr),\notag\allowdisplaybreaks\\
&\quad \textrm{where scalar coefficients $a_{\pm}$ are}\notag\\
&a_{-}=\frac{-2 I \ba \w \cA}{\sqrt{2} \sqrt{\ba\cdot\ba+\cA\cdot\cA+\sqrt{(\ba\cdot\ba+\cA\cdot\cA)^2 -4  (\ba\w\cA)^2}}},\\
&a_{+}=\frac{\sqrt{\ba\cdot\ba+\cA\cdot\cA+\sqrt{(\ba\cdot\ba+\cA\cdot\cA)^2 -4  (\ba\w\cA)^2}}}{\sqrt{2}}\,\notag \\
&\qquad \textrm{when}\quad \ba \w \cA \neq 0\notag, \quad \textrm{and}\allowdisplaybreaks\\
&\begin{cases}
  a_{+}=\sqrt{\ba\cdot\ba+\cA\cdot\cA},\quad a_{-}=0,& \ba\cdot\ba+\cA\cdot\cA > 0\\
  a_{+}=0,\quad a_{-}=\sqrt{-(\ba\cdot\ba+\cA\cdot\cA)},& \ba\cdot\ba+\cA\cdot\cA < 0,\notag
\end{cases}\\
&\qquad \textrm{when}\quad \ba \w \cA = 0\notag.
\end{align}
Since the exponential ~\eqref{exp30free} is in a coordinate-free
form the above formulas are valid for both mutually isomorphic
$\cl{3}{0}$ and $\cl{1}{2}$ algebras.  If one expands the formulas
into coordinates, of course, the resulting expressions will differ
by signs of some terms. Note that determinant\footnote{In 3D
algebras the  determinant of MV $\m{A}$ is defined by
$\Det(\m{A})=\m{A}\cliffordconjugate{\m{A}}\gradeinverse{\m{A}}\reverse{\m{A}}$
\cite{Shirokov2020a,Marchuk2020}.} $\Det(\ba+\cA)=(a_{\smash{+}}^2+a_{\smash{-}}^2)^2$. When
$\Det(\ba+\cA)=0$ we have special case $\exp(\m{A})=\ee^{a_0}(\cos
a_{123}+I\sin a_{123})$ which again can be straightforwardly
obtained by computing limit of~\eqref{exp30free}, when both
$a_{+}\to 0$ and $a_{-}\to 0$. Simultaneous vanishing of $a_{+}$
and $a_{-}$ means vanishing of both the inner $\ba\cdot\ba+\cA\cdot\cA$
and outer $\ba \w \cA$ products. Then the well-known
Moivre-type formulas from Eq.~\eqref{exp30free} are
\begin{align}
  \exp{\m{A}} &=\ee^{a_0}(\cos a_{123}+\sin a_{123} I) & \ba&=\cA =0,\label{speccl30} \\
  \exp{\m{\cA}}  &=\cos\magnitude{\cA}+\frac{\cA}{\magnitude{\cA}}\sin\magnitude{\cA} & a_0&=a_{123}=\ba=0,\\
  \exp{\m{\ba}}   &=\cosh\magnitude{\ba}+\frac{\ba}{\magnitude{\ba}}\sinh\magnitude{\ba} &
  a_0&=a_{123}=\cA=0.
\end{align}
The~equation~\eqref{speccl30}  represents a special case when
$a_{+}= a_{-}=0$.

 Similarly to \cl{0}{3} algebra (see
Subsec.~\ref{entanglement}), in the exponential~\eqref{exp30free}
the vector and bivector may be disentangled if we assume that
$\ba\|\cA$, i.e. the vector $\ba$ lies in the plane $\cA$. Then
vector-bivector sum can be expressed by trigonometric and
hyperbolic functions in the both \cl{3}{0} and \cl{1}{2} algebras,
\begin{equation}\label{entangleCL03}
 \exp{\m{A}}=
 \begin{cases}
 &\cos\sqrt{\magnitude{\cA}^2-\magnitude{\ba}^2}+\frac{\ba+\cA}{\sqrt{\magnitude{\cA}^2-\magnitude{\ba}^2}}\sin\sqrt{\magnitude{\cA}^2-\magnitude{\ba}^2}\qquad \text{if\ }\ba^2<\cA^2, \\
 & \cosh\sqrt{\magnitude{\ba}^2-\magnitude{\cA}^2}+\frac{\ba+\cA}{\sqrt{\magnitude{\ba}^2-\magnitude{\cA}^2}}\sinh\sqrt{\magnitude{\ba}^2-\magnitude{\cA}^2}\quad \text{if\
 }\ba^2>\cA^2.
 \end{cases}
\end{equation}

\vspace{3mm} \textbf{Example 2.} {\it Exponential of  MV in
\cl{3}{0}.} Let's take the same MV $\m{A}=-8-6 \e{2}-9 \e{3}+5
\e{12}-5 \e{13}+6 \e{23}-4 \e{123}$  as in Example~1 and calculate
the exponential using the coordinate-free
expression~\eqref{exp30free}. We find $\ba\cdot\ba+\cA\cdot\cA=31$, $-2
I \ba \w \cA=-150$. Then $a_{-}=-75
\sqrt{\frac{2}{31+\sqrt{23461}}}$ and
$a_{+}=\sqrt{\tfrac{31+\sqrt{23461}}{2}}$. The exact numerical
answer therefore is
\begin{align}\label{example2}
  \exp(\m{A})= &\frac{1}{\ee^8} \bigl(\cos (4)-\sin (4) I\bigr)\biggl(\cos \Bigl(75 \sqrt{\tfrac{2}{31+\sqrt{23461}}}\Bigr) \cosh \Bigl(\sqrt{\tfrac{31+\sqrt{23461}}{2}}\Bigr)\notag\\
  &\hphantom{\frac{1}{\ee^8} \bigl(\cos (4)-\sin (4) I\bigr)\biggl(}
  -\sin \Bigl(75 \sqrt{\tfrac{2}{31+\sqrt{23461}}}\Bigr) \sinh \Bigl(\sqrt{\tfrac{31+\sqrt{23461}}{2}}\Bigr) I\notag\allowdisplaybreaks\\
  &+\frac{1}{\sqrt{23461}}\biggl(\Bigl(-75 \sqrt{\tfrac{2}{31+\sqrt{23461}}} (-6 \e{2}-9 \e{3}+5 \e{12}-5 \e{13}+6 \e{23})\notag\\
  &\hphantom{+\frac{1}{\sqrt{23461}}\biggl(\Bigl(}
  +\sqrt{\tfrac{31+\sqrt{23461}}{2}}  (-6 \e{2}-9 \e{3}+5 \e{12}-5 \e{13}+6 \e{23}) I \Bigr)\notag\\
  &\quad\times\Bigl( -\sin \Bigl(75 \sqrt{\tfrac{2}{31+\sqrt{23461}}}\Bigr) \cosh \Bigl(\sqrt{\tfrac{31+\sqrt{23461}}{2}}\Bigr)\notag \\
  &\quad\qquad-\cos \Bigl(75 \sqrt{\tfrac{2}{31+\sqrt{23461}}}\Bigr) \sinh \Bigl(\sqrt{\tfrac{31+\sqrt{23461}}{2}}\Bigr) I
  \Bigr)\biggr)\biggr).
\end{align}

\vspace{3mm}  \textbf{Example 3.}  {\it Exponential in \cl{1}{2}
with disentanglement included.}  Let's take a simple MV,
$\m{A}=3-\e{1}+2\e{12}$, which represents disentangled case
$\ba\w\cA=0$. Then we have $a_{+}= \sqrt{5}$ and $a_{-}=0$.
The answer is expressed in hyperbolic functions:
$
\exp(\m{A})= \ee^3 \Bigl(\cosh \sqrt{5} +(-\e{1} + 2\e{12})\frac{\sinh \sqrt{5}}{\sqrt{5}}  \Bigr)
$.


\section{MV exponential  in \cl{2}{1}}
\label{sec:Cl21}

After assembling $\cl{2}{1}$ coefficients given in
\cite{Dargys2021} into MV and then regrouping them in a
coordinate-free form we have
\begin{align}\label{exp21free}
\exp(\m{A})=
  &\frac{1}{2}\ee^{a_0}
  \Bigl(
  \ee^{a_{123}} (1 + I)  \bigl( \mathop{\mathrm{co}}(a_{+}^2)+\mathop{\mathrm{si}}(a_{+}^2) (\ba + \cA)
     \bigr)\\
      &\qquad + \ee^{-a_{123}}(1 -I ) \bigl(\mathop{\mathrm{co}}(a_{-}^2)+\mathop{\mathrm{si}}(a_{-}^2)  (\ba + \cA)\bigr)\Bigr)\notag\\
&\quad \textrm{where scalar coefficients $a_{\pm}$ are}\notag\\
  &a_{-}^2=-(\ba\cdot\ba+\cA\cdot\cA) + 2I  \ba\w\cA,\quad a_{-}^2\gtrless 0,\label{apliusminuscl21}\\
  &a_{+}^2=-(\ba\cdot\ba+\cA\cdot\cA) - 2I  \ba\w\cA,\quad a_{+}^2\gtrless 0.\notag\\
\end{align}
To simplify notation  in Eq.~\eqref{exp21free} we have introduced
$\mathop{\mathrm{si}}$ and $\mathop{\mathrm{co}}$ functions that
depending on sign under square root go over to trigonometric or
hyperbolic functions. In $\mathop{\mathrm{si}}$ and
$\mathop{\mathrm{co}}$ functions, either lower or upper signs
should be included. All in all, this gives four cases for both si
and co,
\begin{align} \mathop{\mathrm{si}}(a_{+}^2)=&\begin{cases}
    \frac{\sin \sqrt{a_{+}^2}}{\sqrt{a_{+}^2}},& a_{+}^2 > 0\\
    \frac{\sinh \sqrt{-a_{+}^2}}{\sqrt{-a_{+}^2}},& a_{+}^2 <0 \\
  \end{cases};\qquad \mathop{\mathrm{co}}(a_{+}^2)=\begin{cases}
   \cos \sqrt{a_{+}^2},& a_{+}^2 > 0\\
   \cosh \sqrt{-a_{+}^2},& a_{+}^2 <0 \\
  \end{cases}
\end{align}
\begin{align} \mathop{\mathrm{si}}(a_{-}^2)=&\begin{cases}
    \frac{\sin \sqrt{a_{-}^2}}{\sqrt{a_{-}^2}},& a_{-}^2 > 0\\
    \frac{\sinh \sqrt{-a_{-}^2}}{\sqrt{-a_{-}^2}},& a_{-}^2 <0 \\
  \end{cases};\qquad \mathop{\mathrm{co}}(a_{-}^2)=\begin{cases}
   \cos \sqrt{a_{-}^2},& a_{-}^2 > 0\\
   \cosh \sqrt{-a_{-}^2},& a_{-}^2 <0 \\
  \end{cases}
\end{align}
When $a_{-}=0$ and/or $a_{+}=0$ we have special cases, which again
can be easily included taking already mentioned limits, i.e., by
putting $\mathop{\mathrm{co}}(0)=1$ and
$\mathop{\mathrm{si}}(0)=1$.

\subsection{Special cases}
If  both  the vector $\ba$ and bivector  $\cA$ are equal to zero
the exponential~\eqref{exp21free} simplifies to
$\exp{\m{A}}=\exp(a_0+I a_{123})=\ee^{a_0}(\cosh a_{123}+I \sinh
a_{123})$.

The exponential of vector, when $a_0=a_{123}=\cA=0$, is
\begin{equation}
\exp(\ba)=
\begin{cases}
 &\cos\sqrt{-\ba^2}+\frac{\ba}{\sqrt{-\ba^2}} \sin\sqrt{-\ba^2}\quad\text{if \ } \ba^2<0,\\
 &\cosh\sqrt{\ba^2}+\frac{\ba}{\sqrt{\ba^2}} \sinh\sqrt{\ba^2}\qquad\text{if \ }
 \ba^2>0.
\end{cases}
\end{equation}

The exponential of bivector, when $a_0=a_{123}=\ba=0$, is
\begin{equation}
\exp(\cA)=
\begin{cases}
 &\cos\sqrt{-\cA^2}+\frac{\cA}{\sqrt{-\cA^2}} \sin\sqrt{-\cA^2}\quad\text{if \ } \cA^2<0,\\
 &\cosh\sqrt{\cA^2}+\frac{\cA}{\sqrt{\cA^2}} \sinh\sqrt{\cA^2}\qquad\text{if \ }
 \cA^2>0.
\end{cases}
\end{equation}
If $a_0$ and $a_{123}$ are not equal to zero then $\exp(\ba)$ and
$\exp(\cA)$ should be multiplied by $\ee^{a_0}(\cosh a_{123}+I
\sinh a_{123})$.

 \vspace{3mm} \textbf{Example 4.} {\it
Exponential of  MV in \cl{2}{1}.}\\
 {\it 1.~Case
$a_{-}^2<0,a_{+}^2>0$.} Using the same MV $\m{A}=-8-6 \e{2}-9
\e{3}+5 \e{12}-5 \e{13}+6 \e{23}-4 \e{123}$ for \cl{2}{1} we have
$a_{-}^2=-141$, $a_{+}^2=159$. The answer then is
\begin{align}
  \exp(\m{A})= &\frac{1}{2\ee^8} \biggl(\frac{1}{\ee^4}
  (1+I)\Bigl(\tfrac{\sin \sqrt{159}}{\sqrt{159}} \bigl(-6 \e{2}-9 \e{3}+5 \e{12}-5 \e{13}+6 \e{23}\bigr)\notag\\
  &\hphantom{\frac{1}{2\ee^8} \biggl(\frac{1}{\ee^4} (1+I)\Bigl(\tfrac{\sin \sqrt{159}}{\sqrt{159}} \bigl(-6 \e{2}-9 \e{3}+5 \e{12}-5 \e{13}}
    +\cos \sqrt{159}\Bigr)\notag\\
  & +\ee^4  (1-I)\Bigl(\tfrac{\sinh \sqrt{141}}{\sqrt{141}} (-6 \e{2}-9 \e{3}+5 \e{12}-5 \e{13}+6 \e{23})\notag\\
  &\hphantom{\frac{1}{2\ee^8} \biggl(\frac{1}{\ee^4} (1+I)\Bigl(\tfrac{\sin \sqrt{159}}{\sqrt{159}} \bigl(-6 \e{2}-9 \e{3}+5 \e{12}-5 \e{13}}
  +\cosh \sqrt{141}\Bigr)\biggr)\notag.
\end{align}

\textbf{Example 5.}  {\it Exponential in \cl{2}{1}. Case
$a_{-}^2<0,a_{+}^2<0$.} Exponentiating $\m{A}=-6 \e{2}+5 \e{12}+
\e{123}$ of \cl{2}{1} we have $a_{-}^2=-11$, $a_{+}^2=-11$. The
answer then is 
\begin{align}
  \exp(\m{A})= &\frac{1}{2} \biggl(\bigl(\ee
  (1+I)+\ee^{-1}(1-I)\bigr)\Bigl(\tfrac{\sinh \sqrt{11}}{\sqrt{11}} \bigl(-6 \e{2}+5 \e{12}\bigr) +\cosh \sqrt{11}\Bigr)\biggr).\notag
\end{align}

\vspace{3mm} \textbf{Example 6.} {\it Exponential in \cl{2}{1}.
Case $a_{-}^2>0,a_{+}^2>0$.} Exponentiating
$\m{A}=2+\e{3}+6 \e{12}+ 3\e{123}$ of \cl{2}{1} we have
$a_{-}^2=49$, $a_{+}^2=25$. The answer then is 
\begin{align}
  \exp(\m{A})= &\frac{\ee^2}{2} \Bigl(\ee^3
  (1+I)\bigl(\tfrac{\sin 5}{5} \bigl(\e{3}+ 6\e{12}\bigr) +\cos 5\bigr)\notag\\
  & +\ee^{-3} (1-I)\bigl(\tfrac{\sin 7}{7} (\e{3}+6 \e{12})
  +\cos 7\bigr)\Bigr)\notag.
\end{align}

\vspace{3mm} \textbf{Example 7.}  {\it Exponential in \cl{2}{1}.
Case $a_{-}^2>0,a_{+}^2<0$.} Exponentiating $\m{A}=2-10\e{2}-10
\e{3} +2\e{13}+ \e{23}+\e{123}$ of \cl{2}{1} we have $a_{-}^2=35$,
$a_{+}^2=-45$. The answer then is
\begin{align}
  \exp(\m{A})= &\frac{\ee^2}{2} \Bigl(\ee
  (1+I)\bigl(\tfrac{\sinh 3\sqrt{5}}{3\sqrt{5}} \bigl(-10\e{2}-10 \e{3} +2\e{13}+ \e{23}\bigr) +\cosh(3\sqrt{5})\bigr)\notag\\
  & +\ee^{-1} (1-I)\bigl(\tfrac{\sin \sqrt{35}}{\sqrt{35}} (-10\e{2}-10 \e{3} +2\e{13}+ \e{23})
  +\cos \sqrt{35}\bigr)\Bigr)\notag.
\end{align}

\section{Examples of application: Solution of GA differential equations}
\label{sec:Application}

The exponential function plays a fundamental role in solution of
linear  differential equations. For example, the solution of a
homogeneous  equation for MV $\m{X}$,
\begin{equation}\label{eq1}
\frac{\dd\m{X}(t)}{\dd t}=\m{X}(t),\quad\m{X}(0)=\m{X}_0,
\end{equation}
is GA exponential $\m{X}(t)=\ee^{t\m{A}}\m{X}_0$, where $t$ is the
parameter (in physics usually it is the time). Treating $t\m{A}$
as a new MV, after expansion of exponential we will obtain the
evolution of MV in time. More generally, with suitable assumptions
upon smoothness of $\m{X}(t)$, the solution of the inhomogeneous
system
\begin{equation}\label{eq2}
\frac{\dd\m{X}(t)}{\dd t}=\m{X}(t)+f(t),\quad\m{X}(0)=\m{X}_0,
\end{equation}
may be expressed by
\begin{equation}
\m{X}(t)=\ee^{t\m{A}}\m{X}_0+\int_0^t\ee^{(t-s)\m{A}}f(s)\dd s.
\end{equation}

Some of differential MV equations, for example,
\begin{equation}\label{eq3}
\frac{\dd\m{X}(t)}{\dd t}=\m{A}\m{X}(t)+\m{B}\m{X}(t),\quad
\m{X}(0)=\m{X}_0,
\end{equation}
have  solutions  that  consist  of products  of GA exponentials:
$\m{X}(t)=\ee^{t\m{A}}\,{X}_0\ee^{t\m{B}}$. The answer  can be
easily checked by direct substitution of $\m{X}(t)$ into
Eq.~\eqref{eq3} and application of Leibniz's
differentiation theorem~\cite{Doran03} .

Trigonometric GA functions, as well as GA roots, arise in the
solution of second order differential equations. For example, the
GA equation~\cite{Higham08}
\begin{equation}\label{eq4}
\frac{\dd^2\m{X}(t)}{\dd
t^2}+\m{A}\m{X}=0,\quad\m{X}(0)=\m{X}_0\text{\ and \
}(\dd\m{X}/\dd t)_{t=0}=\m{X}^{\prime}_0
\end{equation}
has the solution
\begin{equation}\label{eq5}
\m{X}(t)=\cos(\sqrt{\m{A}}\,
t)\m{X}_0+\big(\sqrt{\m{A}}\big)^{-1}\sin(\sqrt{\m{A}}\,
t)\m{X}_0^{\prime},
\end{equation}
where $\sqrt{\m{A}}$ is the square root of $\m{A}$. The
trigonometric functions of MV argument can be expressed by
exponentials~\cite{Dargys2021}.  Closed form expression for square
root of MV in $p+q=3$ can be found
in~\cite{AcusDargysPreprint2020}.

 A concrete example of application of GA exponential in physics  can be found
in~\cite{Dargys2021}.

\section{Discussion}
\label{sec:discussion}

The main results are the formulas \eqref{exp03free},
\eqref{exp30free}  and \eqref{exp21free}, where the GA
exponentials are presented in an expanded coordinate-free form.
Since in 3D algebras the scalar and pseudoscalar belong to the GA
center the respective real coefficients $a_0$ and $a_{123}$ appear
in scalar exponentials only. However,  entanglement (mixing) of
vector and bivector components takes place. The mixing  is
characterized by scalar coefficients $a_{+}$ and $a_{-}$, where
the terms of the form $(a_i-a_{jk})^2$, $i\ne j\ne k$ appear. The
disentanglement can be done by equating either vector or bivector
to zero what gives the well-known trigonometric-hyperbolic
Moivre-type formulas for vector and bivector. However, more
interesting is that the disentanglement also can be achieved if
the vector is parallel to bivector as a result we obtain the
exponential which consists of a sum of scalar, vector and
bivector, Eqs.~\eqref{entangleA} and \eqref{entangleCL03}.

The related to exponential is the logarithm function. GA logarithm
problem is more difficult to analyze~\cite{Acus2021b}. Contrary to
exponential the logarithm, similarly to complex logarithm, is a
not a single valued function and  therefore  one must proceed with
caution not to mix different branches. In addition,  one must
introduce principle value

Finally, the GA exponentials  with complex coefficients also need
deeper analysis, more so, since the relativity theory may be
introduced by complexified 3D rather then 4D
algebras~\cite{Baylis99},



%
\bibliographystyle{REPORT}
\bibliography{expdim3}

\end{document}